\documentclass[letterpaper, 10 pt, conference]{ieeeconf}  

\usepackage{caption}
\usepackage{subcaption}
\usepackage{graphicx}
\usepackage{siunitx}
\usepackage{textcomp} 
\usepackage{amsmath,amssymb}
\usepackage{float}
\usepackage{physics}
\usepackage{comment}
\usepackage{textcomp}
\usepackage{xcolor}
\IEEEoverridecommandlockouts                              

\title{\LARGE \bf
Fiber Activation by Bipolar Stimulation in Deep Brain Stimulation:\\ A Patient Case Study*
}

\author{Anna Franziska Frigge$^{1}$, Elena Jiltsova$^{2}$, Fredrik Olsson$^{3}$, Dag Nyholm$^{2}$, and Alexander Medvedev$^{1}$
\thanks{*This work is part of the project "Patient-specific dynamical modeling and optimization of deep brain stimulation" funded within The EU Joint Programme – Neurodegenerative Disease Research 
by the Swedish Research Council, Grant 2020-02901.}
\thanks{$^{1}$ Information Technology,
        Uppsala University, SE-75236 Uppsala, Sweden
        {\tt\small \{anna.frigge,alexander.medvedev\}@it.uu.se}}
\thanks{$^{2}$ Neurology, Uppsala University Hospital, SE-75185 Uppsala, Sweden
        {\tt\small \{elena.jiltsova,dag.nyholm\}@neuro.uu.se}}   
\thanks{$^{3}$ Stardots AB, 75323 Uppsala, Sweden \newline
        {\tt\small fredrik.olsson@stardots.se}} 
        }

\begin{document}
\maketitle
\thispagestyle{empty}
\pagestyle{empty}

\begin{abstract}
Deep Brain Stimulation (DBS) is a therapy widely used for treating the symptoms of neurological disorders. Electrical pulses are chronically delivered in DBS to a disease-specific brain target via a surgically implanted electrode. The stimulating contact configuration, stimulation polarity, as well as amplitude, frequency, and pulse width of the DBS pulse sequence are utilized to optimize the therapeutic effect. In this paper, the utility of  therapy individualization by means of patient-specific mathematical modeling is investigated  with respect to a specific case of a patient diagnosed with Essential Tremor (ET). 
Two computational models are compared in their ability to elucidate the impact of DBS stimulation on the dentato-rubrothalamic tract: (i) a conventional model of Volume of Tissue Activated (VTA) and (ii) a well-established neural fiber activation modeling framework known as OSS-DBS. The simulation results are compared with tremor measured in the patient under different DBS settings using a smartphone application.
The findings of the study highlight that temporally static VTA models do not adequately describe the differences in the outcomes of bipolar stimulation settings with switched polarity, whereas neural fiber activation models hold potential in this regard. However, it is noted that neither of the investigated models fully accounts for the measured symptom pattern, particularly regarding   a bilateral effect produced by unilateral stimulation.

\end{abstract}

\section{Introduction}
Deep Brain Stimulation (DBS) is a well-established medical procedure that was approved by the Food and Drug Administration for the treatment of Parkinson's Disease (PD) and Essential Tremor (ET) in 1997. While its therapeutic efficacy is widely recognized, there is an on-going debate over the most suitable brain targets for achieving optimal symptom relief. Recent studies point towards the stimulation of specific fiber tracts as effective treatment targets in the alleviation of symptoms in ET~\cite{Middlebrooks2021,Deuter2022} and PD~\cite{Butenko2022}.
This paper focuses on modeling the activation of the Dentato-Rubro-Thalamic tract (DRTT), which has been identified as a potential pathway for symptom alleviation in ET~\cite{Middlebrooks2021}.

\begin{figure}
      \centering
      \includegraphics[angle=0,origin=c,scale=0.105]{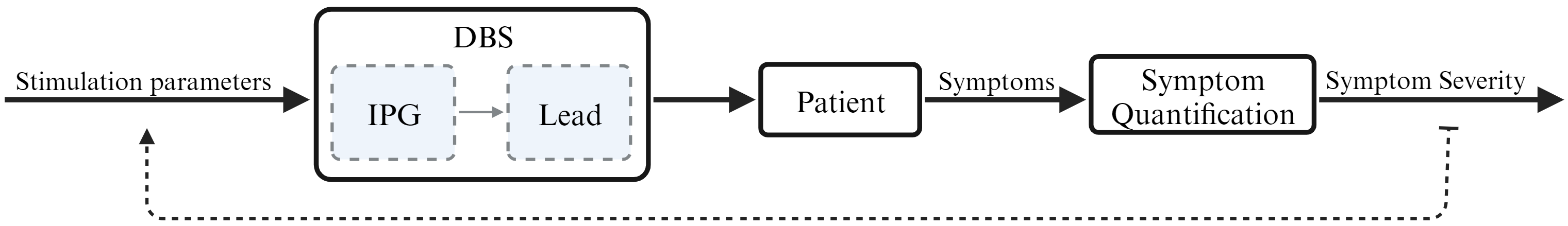}
      \caption{DBS as a control system. The DBS system comprises the implanted pulse generator (IPG), the lead, and the connecting wire; the stimulation parameters serve as input. Patient symptoms are quantified for symptom severity assessment.  The quantified symptoms are fed back to tune stimulation parameters in closed-loop DBS.}
      \label{fig:control_system}
\end{figure}

DBS can be viewed as an implantable control system, see Fig.~\ref{fig:control_system}, where the stimulation parameters act as inputs to the patient (plant), and observed symptom severity constitutes the controlled output~\cite{Medvedev2019}. Currently, the control is performed in an open-loop manner with fixed parameters of the stimulation signal but the development of closed-loop DBS systems, both using signals measured in the brain and obtained by symptom quantification, are under way~\cite{Cuschieri2022}.

Despite the well-documented therapeutic efficacy of DBS, the exact neural mechanisms underlying  the effects of stimulation remain unclear. Currently, there are three main theories under discussion regarding how the stimulation impacts the neural structures of the target: (i) the inhibition of neural signals, (ii) the excitation of neural signals, and (iii) the disruption of abnormal information flow~\cite{Chiken2016}. This ambiguity significantly complicates  computational modeling and  the development of model-based therapy optimization in DBS.

Modern implanted pulse generators (IPG) for DBS are able to produce both unipolar and bipolar stimulation.
In unipolar settings, the implanted pulse generator serves as the anode. Conventionally, a unipolar review, involving the examination of each contact with cathodic stimulation, is employed to identify the most effective contact configuration. Due to the high number of possible combinations, bipolar settings, where one (or multiple) contacts are designated as cathode, while other contacts serve as the anode, are typically not explored until all unipolar options have been exhaustively tested and have failed to produce the desired symptom relief. In the advanced DBS leads intended for stimulation field steering~\cite{Toader2010} via numerous segmented contacts, unipolar review becomes infeasible due to the combinatorial nature of the problem. 

The stimulation’s alleviating effect on  tremor is typically occuring on the side opposite to the stimulation (contralateral), nevertheless, in some patients an effect on the same
side (ipsilateral) can also be observed. The mechanisms underlying the influence of unilateral stimulation on both sides and the interplay between stimulation in both hemispheres remain poorly understood, which  poses another challenge in the search for optimal stimulation settings~\cite{Hasegawa2020}.

Optimization strategies for identifying the most effective contacts for individual patients usually neglect bipolar configurations as well as the interplay between ipsi- and contralateral stimulation effect, primarily because viable methods for assessing their impact are currently missing.

The contributions of this paper are as follows:
\begin{enumerate}
    \item Simulation of bilateral therapeutic response i.e. stimulation in one hemisphere that  results in tremor reduction on both sides. 
    \item Utility comparison of two patient-specific computational DBS models with respect to fiber activation under bipolar settings.
\end{enumerate}

The subsequent sections will introduce and assess the two model-based methods in terms of their effectiveness in discriminating between bipolar stimulation configurations. 
To our knowledge, this study represents the first attempt to align outcomes derived from personalized computational models of fiber activation in DBS with objective symptom quantification. Specifically, the  presented approach enables individual impact assessment of both unilateral and bilateral stimulation in each hand individually. Moreover, it emphasizes the limitations of temporally static models in capturing the effects of bipolar stimulations with altered polarities, thereby addressing the question of necessary model complexity. 

\section{Methodology}
\subsection{Symptom Control in Essential Tremor}
\begin{figure}
    \centering
    \includegraphics[angle=0,width=\linewidth]{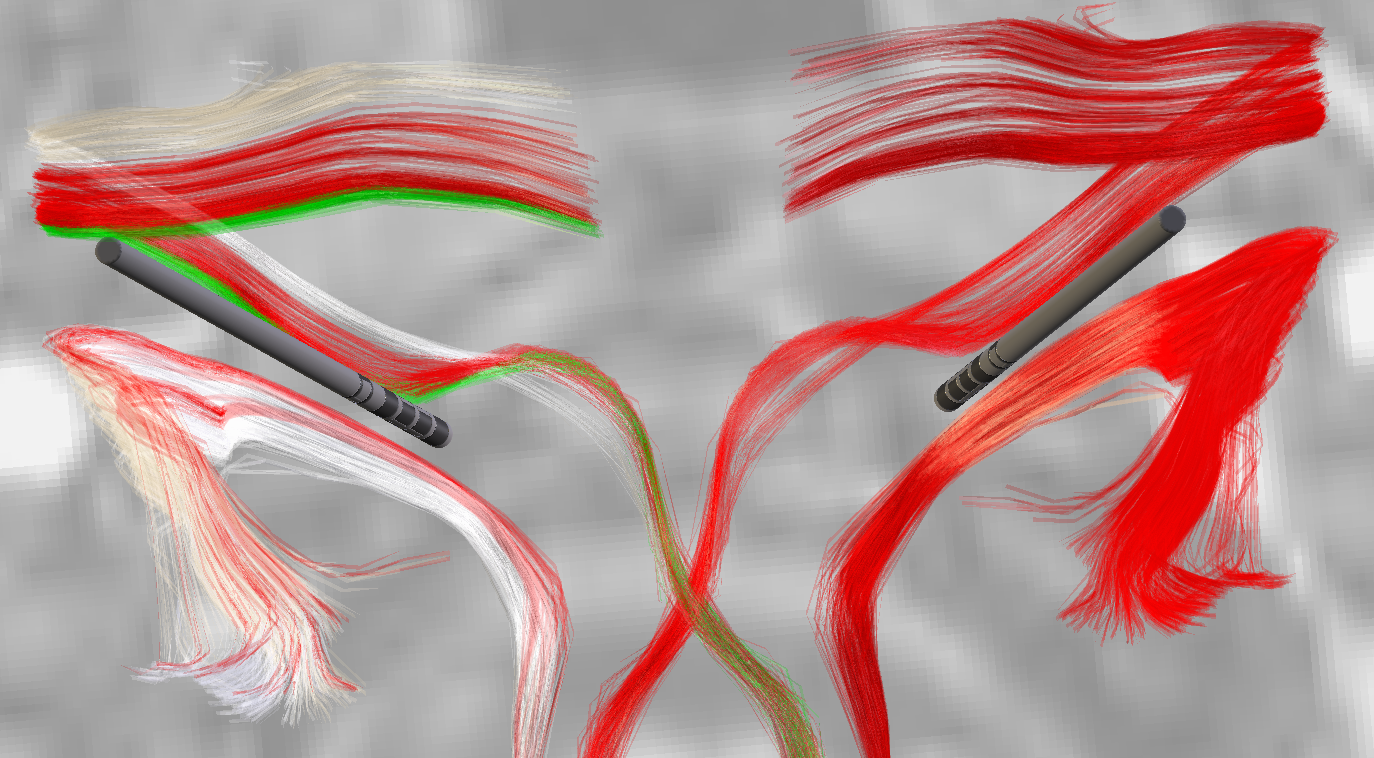}
    \caption{Fiber activation for clinically active settings as given by OSS-DBS. Neural fibers intersecting the lead trajectory or encapsulation layer are labeled as damaged (green), whereas neuron models exhibiting induced firing are categorized as activated (red). Fibers not exhibiting activation are labeled as non-activated (white).}
    \label{fig:Fiber_activation}
\end{figure}

The choice of the stimulation target in DBS is typically guided by the patient's specific symptoms. In the context of Essential Tremor, the ventralis intermedius nucleus  (Vim) is often regarded as the primary target, but alternative targets such as the caudal zona incerta (cZI) and the subthalamic nucleus have also proven efficient~\cite{Chandra2022, Holslag2018}. The heterogeneity of stimulation targets in ET is a topic of on-going debate, but it might be explainable by a shared neural treatment pathway, the Dentato-Rubro-Thalamic Tract (DRTT)~\cite{Middlebrooks2021,Yang2020,Dembek2020}.
The DRTT comprises a larger decussating (dDRTT) and a smaller non-decussating portion (ndDRTT) as shown in Fig.~\ref{fig:Fiber_activation}. While both fibers are associated with a good clinical response in ET, it appears that the dDRTT assumes a more crucial role~\cite{Deuter2022}. Moreover, Middlebrooks \textit{et al.}~\cite{Middlebrooks2021} demonstrated that several previously identified "sweet spots" for stimulation in ET overlap with the DRTT. 

The patient selected for this study\footnote{The study was approved by the Swedish Ethics Review Authority, Dnr 2019–05718.} was diagnosed with ET and DBS was proposed as a treatment option for the severe hand tremor on both sides. Notably, the patient is left-handed, and it is observed that the tremor is more pronounced in the left hand, consistent with the heightened manifestation often seen in the dominant hand. The caudal zona incerta (cZI) was selected as the stimulation target for lead placement. The patient underwent bilateral DBS surgery with the short Abbott Medical Infinity\texttrademark \ directional lead.
After implantation and tuning of stimulation parameters, both the patient and medical staff reported that stimulation in one hemisphere has a therapeutic effect not only on the contralateral side, but also on the ipsilateral side. 

\subsection{DBS Lead and Tissue Modeling}
\begin{figure}
\begin{minipage}{0.5\linewidth}
  \begin{subfigure}{\linewidth}
    \centering
      \includegraphics[width=0.4\textwidth]{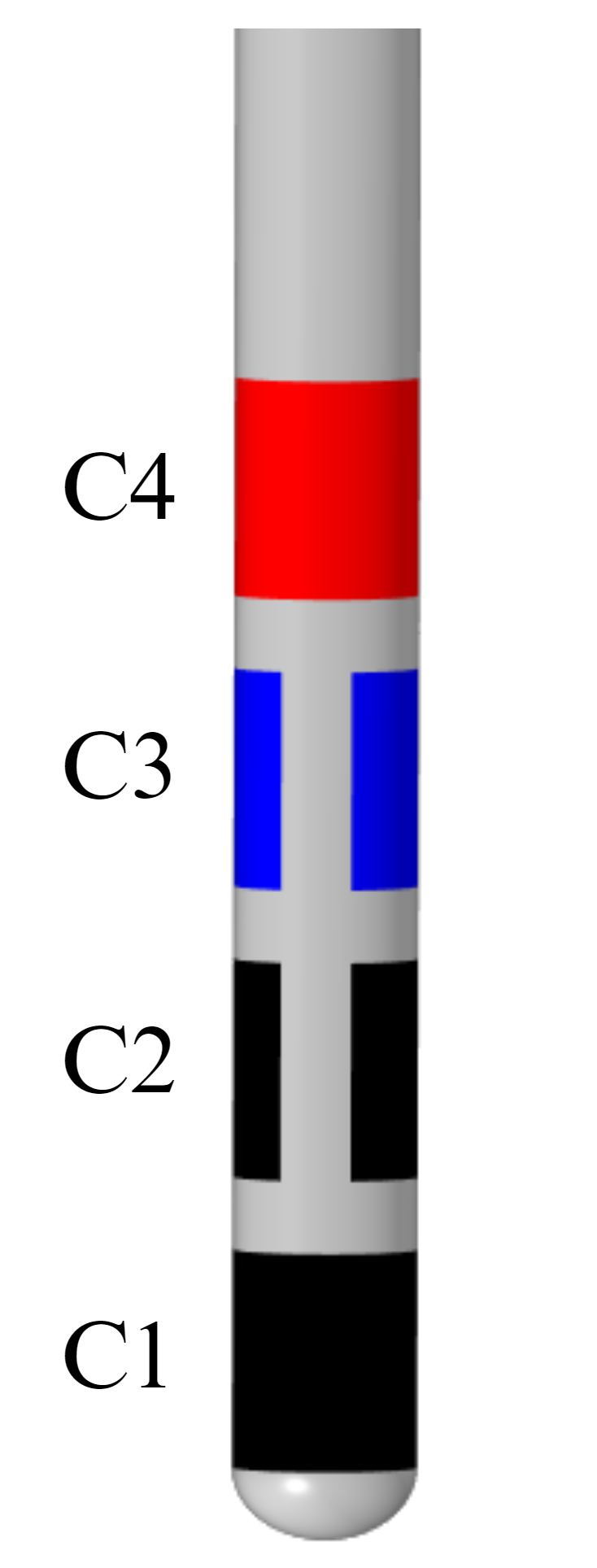}
      \caption{}
      \label{fig:lead_design}
  \end{subfigure}
  
\end{minipage}%
\begin{minipage}{0.5\linewidth}
    \centering
  \begin{subfigure}{\linewidth}
    \centering
    \includegraphics[angle=0,origin=c,width=0.9\textwidth]{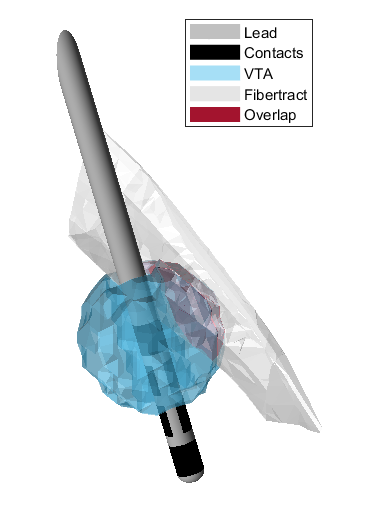}
    \caption{}
  \end{subfigure}
\end{minipage}
      \caption{(a) Visualization of a bipolar C3-, C4+ configuration for the lead. Cathodic contacts are highlighted in blue, and the anodic contact is represented in red. (b) DBS Lead placement relative to a segment of the fibertract, in this case, the ndDRTT. The blue shape represents the static VTA, while the intersection with the fiber tract is shown in red. }
      \label{fig:lead_and_tracts}
\end{figure}

All models discussed in this paper rely on finite element method (FEM) representations of the DBS lead and the adjacent tissue. 
The lead was modeled according to the technical specifications of the short Abbott Medical Infinity\texttrademark \ directional lead, which is depicted in Fig.~\ref{fig:lead_design}. The positioning of the lead is determined from post-operative CT scans. The tissue surrounding the lead is typically simulated either with homogeneous conductivity using a constant value or with spatially varying, heterogeneous conductivity values for different tissue types obtained from MRI data. While most models assume isotropic tissue, certain models incorporate anisotropy by including diffusion tensor imaging (DTI) data from an atlas~\cite{Butenko2020}. In this study, heterogeneous tissue was modeled in a \SI{5}{cm}~$\times$~\SI{5}{cm}~$\times$~\SI{5}{cm} box and by including a \SI{0.1}{mm} encapsulation layer around the lead, neglecting anisotropic effect. Beyond that volume homogeneous dielectric properties were assumed. Values for the dielectric properties are given in Table~\ref{tab:dielectric}. 

\begin{table}[]
    \centering
    \begin{tabular}{c|c|c|c|c|c}
        \multicolumn{6}{c}{\textbf{Dielectric properties}} \\
        \hline
         Tissue type & GM & WM & CSF & Encap. layer & Homog.\\
        \hline
        $\sigma$ [S/m]   & 0.09 & 0.06 & 2.0 & 0.05 & 0.1\\
        $\epsilon_r \cdot 10^4$  & 30.407 & 13.752 & 0.0109 & 30.407 & 13.800
    \end{tabular}
    \caption{Dielectric properties for the different tissue types gray matter (GM), white matter (WM), cerebrospinal fluid (CSF), the encapsulation layer and the homogeneous medium, adopted from~\cite{Butenko2020}. Conductivity values are denoted by $\sigma$, while permittivity values are given by $\epsilon_r$.}
    \label{tab:dielectric}
\end{table}

In time-varying simulations, the electro-quasi static (EQS) approximation of Maxwell's equations, which neglects magnetic induction and is given by

\begin{equation}
    \nabla \cdot \mathbf{J} = -\nabla \cdot (\sigma \nabla u+\epsilon_0 \epsilon_r \nabla u) =0, 
    \label{eq:pde}
\end{equation}

is employed. In this context, $\nabla \cdot$ is the divergence operator, $\mathbf{J}$ represents the current density, $\sigma$ denotes the conductivity, and $\nabla u$ refers to the gradient of the electric potential. The constants $\epsilon_0$ and $\epsilon_r$ denote vacuum permittivity and relative permittivity, respectively.

The EQS approximation given in \eqref{eq:pde} can be simplified to the quasi-static (QS) approximation, 
\begin{equation}
    \nabla \cdot (\sigma \nabla u) = 0,
    \label{eq:pde_static}
\end{equation}
which, with $I(t)=\text{const.}$, omits dispersive effects of the tissue. 

\subsection{Fiber Activation Models}

\paragraph{Static VTA models} 
A common approach for quantifying the static volume of tissue activated (VTA) in Deep Brain Stimulation (DBS) involves the use of an activating function. Among various activating functions, the most simple yet most commonly used method defines the VTA as the volume where the electric field norm generated by the DBS lead exceeds a particular threshold, typically set at 200 V/m. 
However, dynamic stimulation parameters can be incorporated into static VTA models to a certain extent by modifying the VTA threshold. Åström \textit{et al.}~\cite{Astrom2015} provide VTA thresholds from a neuron model for various pulse widths and fiber diameters. 

In this study, the static VTA thresholds were adjusted to match the clinically applied pulse width and a fiber diameter of $\SI{3.5}{\mu m}$. To determine the number of activated fibers in both the dDRTT and the ndDRTT, the magnitude of the electric field was evaluated at atlas coordinates along each tract. Once the electric field exceeded the VTA threshold at a specific coordinate along the fiber tract, the corresponding fiber was marked as activated. The proportion of activated fibers in each tract was calculated by dividing the number of activated fibers by the total number of fibers within that tract.

\paragraph{Neuron models}
The use of neuron models is computationally more expensive compared to the use of temporally static models, however,  they are essential for capturing the highly non-linear, dynamical effects of DBS on neural tissue.
    
In this study, the fiber activation framework OSS-DBS, developed by Butenko \textit{et al.}~\cite{Butenko2020} and distributed with lead-dbs version 3.0~\cite{Neudorfer2023}, is employed. Within the OSS-DBS framework, neural pathway activation under DBS is achieved by incorporating neuron models at specific fiber coordinates. Neuron models that exhibit action potentials or firing under DBS are designated as "activated." The OSS-DBS environment offers a choice between the neuron model developed by McIntyre \textit{et al.} and the neuron model by Reilly \textit{et al.}, both of which are well-recognized in the DBS community. For this paper, the focus was placed on the multi-compartment double cable model for mammalian nerve fibers from McIntyre \textit{et al.}, which explicitly models nodes of Ranvier and internodes. A fiber diameter of $ \SI{5.7}{\mu m}$ was assumed. Further details on this model can be found in~\cite{McIntyre2002}. In Fig.~\ref{fig:Fiber_activation} the DBS leads are portrayed together with activated and non-activated fibers of the DRTT under clinically active settings.

\subsection{Symptom quantification}
\begin{figure}
    \centering
    \includegraphics[width=0.5\linewidth]{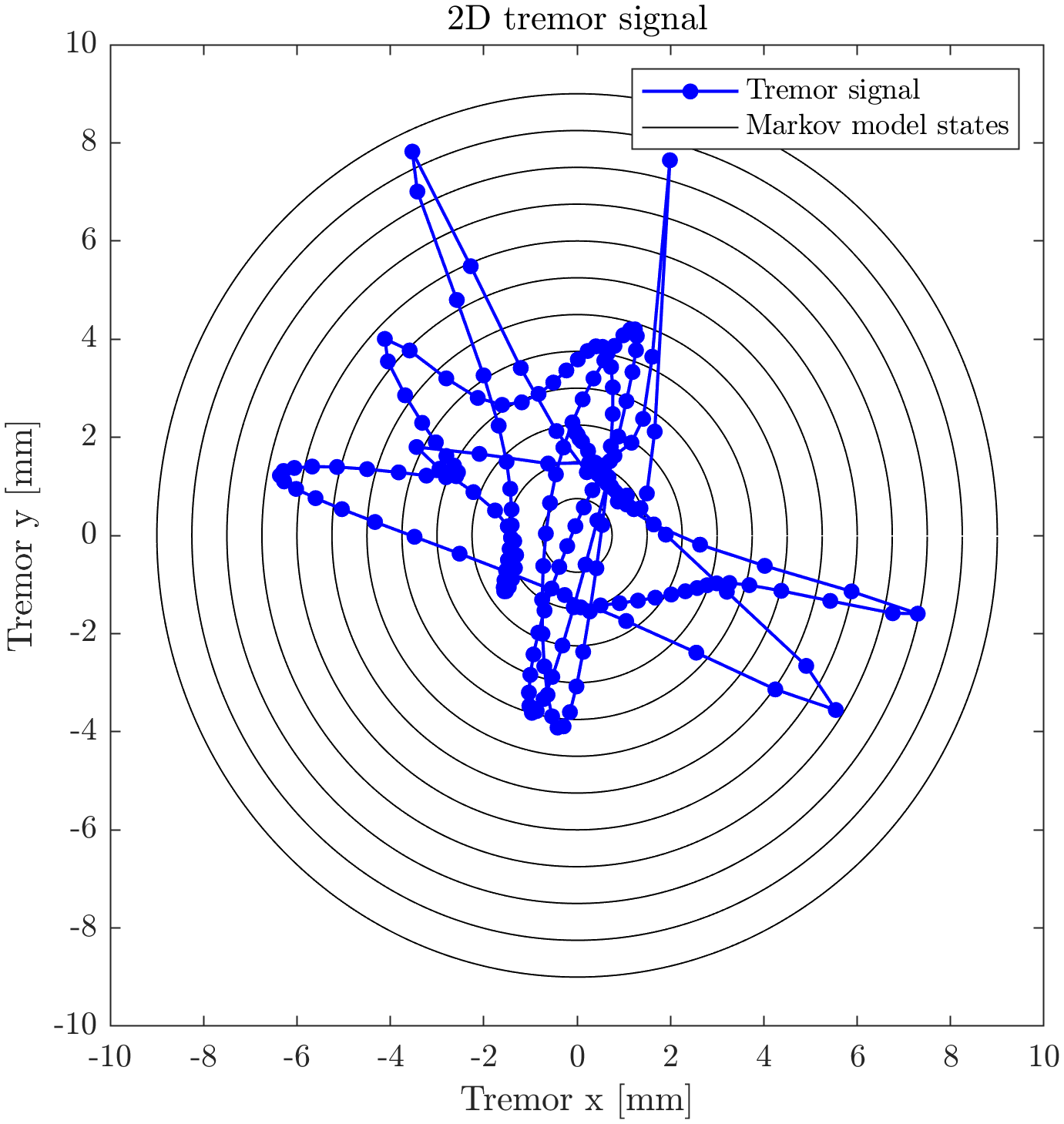}
    \caption{Part of the 2D tremor signal trajectory (blue) under clinical DBS settings for the left hand. Black circles correspond to the upper bound of the radius of the predefined Markov states. }
    \label{fig:markov}
\end{figure}

\begin{figure}
    \centering
    \includegraphics[width=0.43\linewidth]{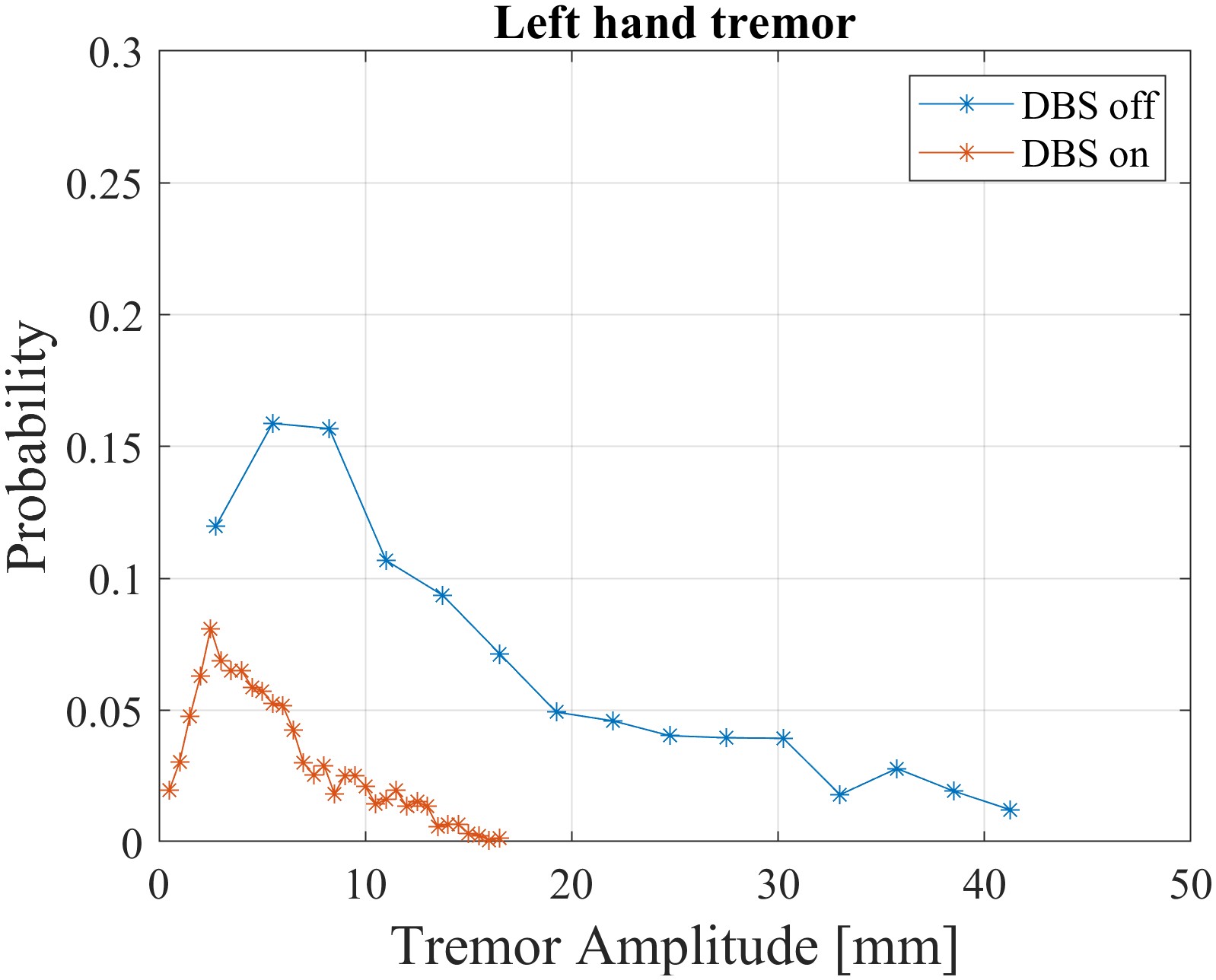} \hspace{0.5cm}
    \includegraphics[width=0.43\linewidth]{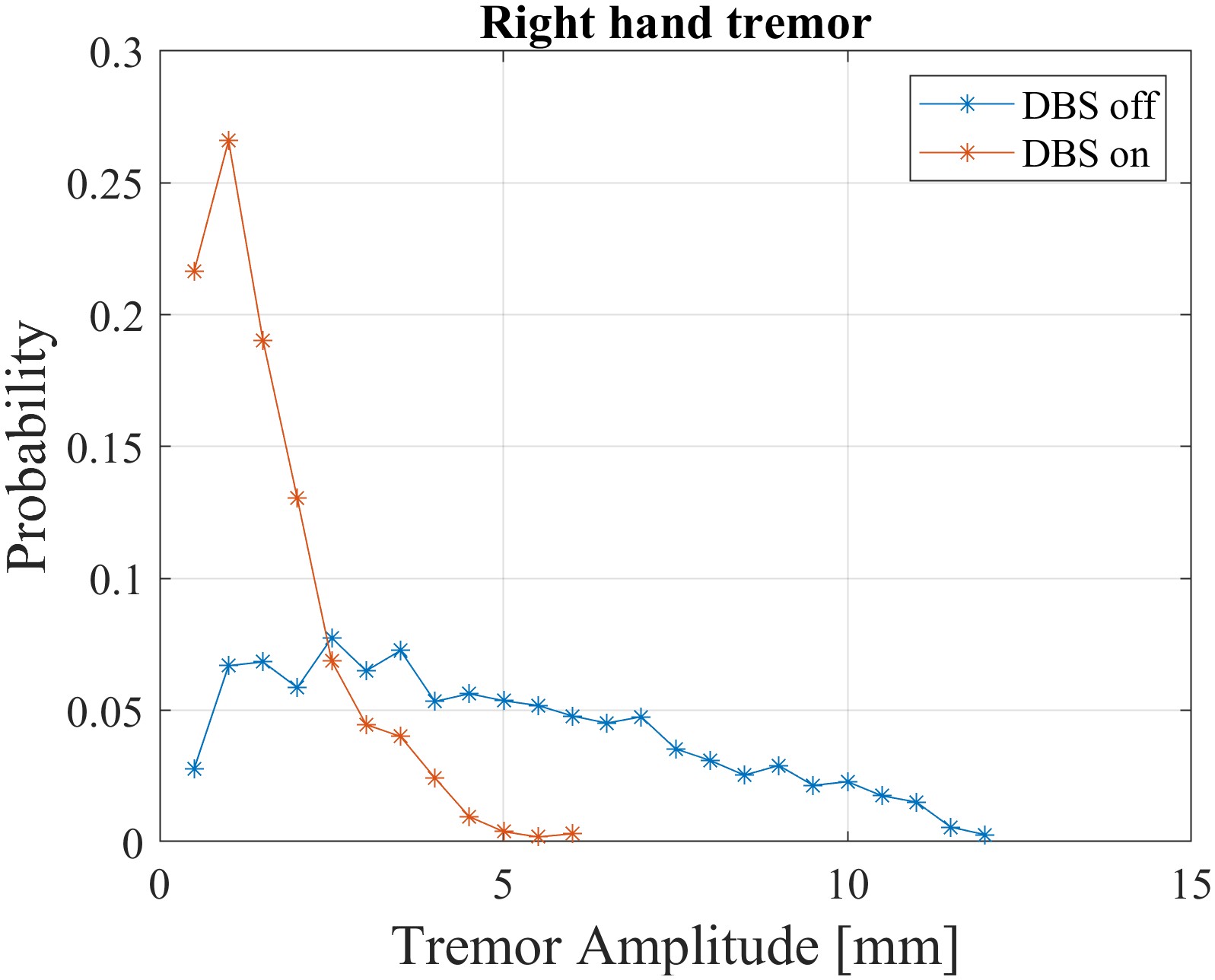}
    \caption{Hand tremor distributions with DBS on and off. The on setting corresponds to the clinically used setting with both leads active. In both hands, the probability of large tremor amplitudes is significantly reduced under DBS. Tremor is more pronounced in the dominant left hand.}
    \label{fig:tremor_distribution}
\end{figure}

In clinical settings, healthcare professionals visually assess patients' symptoms and quantify the results by means of rating scales.  These ratings can be approximately translated to specific ranges of tremor amplitudes. Patients diagnosed with ET are commonly assessed using the Essential Tremor Rating Assessment Scale (TETRAS)~\cite{Elble2012}. An evaluation with TETRAS requires only a pen and paper, and can be completed in about 10~min. Yet, it is predominately  subjective. Furthermore, there have been recent developments in entirely objective quantification methods, using smartphone or smartwatch sensors, which offer an easy and unbiased approach to measuring motor symptoms~\cite{Olsson2020}. Additionally, oculomotor  performance that reflects both motor and non-motor symptoms  can be assessed through the use of eye-tracking~\cite{Ellmerer2022}. In this paper, the smartphone application ANLIVA\textsuperscript{\textregistered}\footnote{ANLIVA is a registered trademark of Stardots AB (www.stardots.se). All rights reserved. The ANLIVA® Hand Movement device is approved for clinical trials in Sweden but is not regulatory approved (CE marked) as a medical device for the European market.} Hand Movement from Stardots AB was applied to evaluate the symptom severity. Tremor amplitudes were estimated from the measurements of the smartphone's inertial sensor platform. During each recording, the patient was instructed to lift the smartphone from a table three consecutive times. The quantification algorithm utilizes a Markov chain model to characterize the tremor amplitude, with states corresponding to predefined severity intervals, as outlined in~\cite{Olsson2020} and illustrated in Fig.~\ref{fig:markov}. Measurements were performed individually for each hand, allowing for an independent tremor assessment for each side. The resulting tremor distributions for the patient with DBS on and off are illustrated in Fig.~\ref{fig:tremor_distribution}. Details on the computation of tremor scores are proprietary of the ANLIVA\textsuperscript{\textregistered} Hand Movement (v.1.3) device (Stardots AB).

\section{Results}
\subsection{Patient measurements}
In Table~\ref{tab:tremor_scores}, the results of the tremor measurements are reported.    	
\begin{table}[]
    \centering
    \begin{tabular}{c|c|c|c|c}
        \multicolumn{4}{c}{\textbf{Tremor scores}} \\
        \hline
        LH Contacts & RH Contacts & LH Tremor& RH Tremor & Total\\
        \hline
        None   & None & 61.9 & 19.7 & 81.6\\
        C3-, C4+  & None & 40.7 & 7.5 & 48.2\\
        None & C2-, C4+  & 20.9 & 12.4 & 33.3\\
         C3-, C4+ & C2-, C4+  & 24.4 & 7.1 & 31.5\\
         C3-, C4+ & C4-, C2+*  & 32.4 & 6.9 & 39.3\\
    \end{tabular}

    \vspace{10pt}
      
    \begin{tabular}{c|c|c}
        \multicolumn{3}{c}{\textbf{Constant Parameters}} \\
        \hline
         Stimulation Paramters & Left Hemisphere & Right Hemisphere\\
        \hline
        Amplitude [\SI{}{mA}] & 1.2  & 4 and 1.6\\
        Frequency [\SI{}{Hz}]  & 140 & 184\\
        Pulse width [\SI{}{\mu s}]  & 90 & 50\\
        Static threshold [\SI{}{V/m}] & 150 & 230\\
    \end{tabular} 
    \caption{Top: Evaluation of tremor scores for the patient-specific clinically used settings was conducted for both the left hand (LH) and right hand (RH) individually, with lower scores signifying reduced tremor severity. The total tremor score represents the sum of the two individual tremor scores. The stimulation's impact on alleviating tremor is more pronounced on the side opposite to the stimulation, but an effect on the same side can also be observed. *Here, the stimulation amplitude was set to \SI{1.6}{mA} instead of \SI{4}{mA} due to adverse side effects observed at higher amplitudes. Bottom: Stimulation parameters that have been kept constant on each side. The VTA threshold for the static model was adjusted with regard to pulse widths in accordance with~\cite{Astrom2015}.}
    \label{tab:tremor_scores}
\end{table}
Evidently, and as intended, stimulation in one hemisphere yields lower tremor scores, thus reduced tremor, in the contralateral hand.  While the effect is more pronounced on the contralateral side, unilateral stimulation also has a tremor-reducing effect on the ipsilateral side in this patient. Furthermore, the lowest total tremor score in both hands was reported when using the clinically active setting, involving bipolar stimulation in both hemispheres. This configuration was preferred by the patient, as it resulted in a highly functional right hand and a substantial reduction in tremor in the left hand.
In the setting, where polarity was switched compared to the clinically active settings, side effects were observed even at a much lower amplitude. Nevertheless, the reduction in tremor at the highest tolerable amplitude was less pronounced in the contralateral hand, making this setting less favourable for the patient.

\subsection{Activation results}
The results of fiber activation based on the static VTA model are given in Table~\ref{tab:static}.
\begin{table}[thpb]
    \centering
    \begin{tabular}{c|c|c}
        \multicolumn{3}{c}{\textbf{Left Hemisphere}} \\
        \hline
        Active Contacts & Activation dDRTT [\%] & Activation ndDRTT [\%]\\
        \hline
        \textbf{C3-, C4+} & \textbf{82.9}  & \textbf{21.4}\\
        C4-, C3+  & 82.9 & 21.4\\
    \end{tabular}
    
    \vspace{10pt}  
    
    \begin{tabular}{c|c|c}
        \multicolumn{3}{c}{\textbf{Right Hemisphere}} \\
        \hline
        Active Contacts & Activation dDRTT [\%] & Activation ndDRTT [\%]\\
        \hline
        \textbf{C2-, C4+} & \textbf{83.4}  & \textbf{82.1}\\
        C4-, C2+  & 83.4 & 82.1\\
        C4-, C2+*   & 70.8 & 27.8\\
    \end{tabular} 

    \caption{Static VTA fiber activation of the dDRTT and ndDRTT for the left and right hemispheres, respectively. The configurations highlighted in bold font represent the clinically used settings. The VTA threshold was adjusted based on the clinically used pulse widths as given in Table~\ref{tab:tremor_scores}.*Here, the stimulation amplitude was set to 1.6 mA instead of 4 mA to match the measurements. }
    \label{tab:static}
\end{table}
All settings resulted in a substantial coverage of the dDRTT. Notably, the coverage of the ndDRTT was significantly higher for most settings in the right hemisphere. As expected, no difference in coverage can be observed between settings with switched polarity.

In Table~\ref{tab:OSS_EQS} and Table~\ref{tab:OSS_QS} the results of the fiber activation model based on OSS-DBS are given for the EQS and the QS formulation, respectively. The activation percentages for both the dDRTT and the ndDRTT are presented for five different settings.
\begin{table}[thpb]
    \centering
    \begin{tabular}{c|c|c}
        \multicolumn{3}{c}{\textbf{Left Hemisphere}} \\
        \hline
        Active Contacts & Activation dDRTT [\%] & Activation ndDRTT [\%]\\
        \hline
        \textbf{C3-, C4+} & \textbf{49.53}  & \textbf{11.07}\\
        C4-, C3+  & 49.3 & 13.79\\
    \end{tabular}
    
    \vspace{10pt}  
    
    \begin{tabular}{c|c|c}
        \multicolumn{3}{c}{\textbf{Right Hemisphere}} \\
        \hline
        Active Contacts & Activation dDRTT [\%] & Activation ndDRTT [\%]\\
        \hline
        \textbf{C2-, C4+} & \textbf{100}  & \textbf{100}\\
        C4-, C2+  & 87.38 & 34.95\\
        C4-, C2+* & 76.64 & 15.53
    \end{tabular} 
    \caption{OSS-DBS fiber activation of the dDRTT and ndDRTT obtained from EQS solution for the left and right hemispheres, respectively. The configurations highlighted in bold font represent the clinically active settings. The remaining stimulation parameters are based on the patient's clinical settings and were kept constant for each side, as given in Table~\ref{tab:tremor_scores}. *Here, the stimulation amplitude was set to 1.6 mA instead of 4 mA to match the measurements.}
    \label{tab:OSS_EQS}
\end{table}

\begin{table}[thpb]
    \centering
    \begin{tabular}{c|c|c}
        \multicolumn{3}{c}{\textbf{Left Hemisphere}} \\
        \hline
        Active Contacts & Activation dDRTT [\%] & Activation ndDRTT [\%]\\
        \hline
        \textbf{C3-, C4+} & \textbf{49.53}  & \textbf{4.17}\\
        C4-, C3+  & 49.3 & 7.54\\
    \end{tabular}
    
    \vspace{10pt}  
    
    \begin{tabular}{c|c|c}
        \multicolumn{3}{c}{\textbf{Right Hemisphere}} \\
        \hline
        Active Contacts & Activation dDRTT [\%] & Activation ndDRTT [\%]\\
        \hline
        \textbf{C2-, C4+} & \textbf{100}  & \textbf{100}\\
        C4-, C2+  & 91.59 & 34.95\\
        C4-, C2+* & 77.1& 14.76
    \end{tabular} 
    \caption{OSS-DBS fiber activation of the dDRTT and ndDRTT obtained from QS solution for the left and right hemispheres, respectively. The configurations highlighted in bold font represent the clinically active settings. The remaining stimulation parameters are based on the patient's clinical settings and were kept constant for each side, as given in Table~\ref{tab:tremor_scores}. *Here, the stimulation amplitude was set to 1.6 mA instead of 4 mA to match the measurements.}
    \label{tab:OSS_QS}
\end{table}
In parallel with the activation results obtained from the static model, a substantial portion of the dDRTT was activated in both hemispheres, with a higher percentage of activation observed in the right hemisphere. While the activation of the ndDRTT was relatively limited in the left hemisphere, all stimulation configurations achieved substantial coverage of the ndDRTT in the right hemisphere. In contrast to the results from the static model, differences in ndDRTT activation were apparent between configurations with reversed polarity in the right hemisphere, whereas minimal differences were observed in the left hemisphere.

\section{Discussion}
In this paper, two fiber activation models were employed to assess the activation of the DRTT and its correlation with observed motor symptoms. The presented symptom quantification approach allows for objective and independent evaluation of tremor on each side. The results confirm the bilateral effect from unilateral stimulation. Additionally, altering the polarity led to a significantly reduced amplitude, which was tolerable compared to the initial settings, emphasizing the crucial role of polarity in DBS programming. 
Temporally static models have proven valuable for visualization and simplification purposes. However, they lack the capability of capturing the dynamic aspects of DBS, particularly when applied to bipolar stimulation configurations. Yet, modeling the effects of bipolar DBS configurations is of particular significance, as these have exhibited comparable or even superior effectiveness to unipolar settings, as illustrated by the patient in this study. Additionally, they hold the potential for advantages like reduced side effects~\cite{Steffen2020} and extended battery life~\cite{Almeida2016}.

Both the static model and the neuron model indicate a higher percentage of activated dDRTT fibers compared to the percentage of activated ndDRTT fibers at the clinically used settings. This suggests that less activation of ndDRTT fibers may be beneficial and that excessive activation of these fibers could potentially lead to side effects. However, it is noteworthy that these clinical settings may not necessarily represent the optimal stimulation parameters. While the right hand tremor is satisfactorily reduced by these settings, the tremor in the dominant left hand remains significant even at the most effective clinical setting identified, as evidenced by the objective tremor score.

The more advanced model including neuron modeling also reports much higher activation of the dDRTT compared to the activation of the ndDRTT in the left hemisphere. In the right hemisphere, both parts of the DRTT are equally activated at clinical settings, yet for switched polarity, the activation of the ndDRTT is reduced significantly. Thus, including neuron models holds potential for distinguishing the effect of different bipolar settings more effectively than static VTA models. For the settings tested, the QS and the EQS method yielded similar activation results.

Notably, neither one of the models considered in this study can clearly account for the interactions between stimulation in one hemisphere and its effects on the measured tremor. The findings presented in this paper are inconclusive regarding the mechanism through which unilateral stimulation leads to tremor reduction on both sides. It raises the question how the dynamic interaction between stimulation in each hemisphere can be described, and whether its effect is associated with the non-decussating and decussating segments of the DRTT. Nevertheless, the presented setup holds potential for exploring how the impact of unilateral stimulation accumulates in bilateral stimulations. In a future study, additional measurements could be employed to derive a function that characterizes the interaction.

The simulations carried out in this study have several limitations. 
The coordinates of the DRTT fiber tract were derived from an atlas because there was no available data for constructing individual pathways for this specific patient. Nevertheless, research by Wang \textit{et al.}~\cite{Wang2021} has shown that both patient-specific and normative tractography tend to yield similar findings regarding the brain regions associated with clinical improvements in DBS. Only the two parts of the DRTT were considered in this study to reduce computation time, yet other tracts that are located in close vicinity to the lead may also contribute to the therapeutic effect.

Furthermore, no traversing activity along the fibers was taken into account. While some argue that traversing activity relative to the DBS pulses is small and therefore negligible~\cite{Butenko2020}, others support the concept of a complex modulation of traversing currents by the DBS pulses, whether through augmentation, suppression, or interruption~\cite{Andersson2018}.

Lastly, the activation models presented in this paper do not take into consideration networks of neurons, but solely consider isolated neurons without interaction. Thus, even though the incorporation of neuron models increases the model complexity tremendously, the problem is still not addressed to full extent. This may or not be necessary to capture the effect of DBS on symptom management completely. Whether a higher level of complexity is necessary to comprehensively capture the impact of DBS on symptom management remains a subject for further investigation.

\section{Conclusions}
The findings outlined in this paper represent a preliminary investigation and do not assert generalizability. Further measurements under a comparable set-up are necessary to explore the bilateral impact of unilateral stimulation and the influence of polarity. Such a study would provide greater insight into the necessary model setup and complexity for accurately correlating simulation outcomes with objective symptom measurements. As demonstrated in this paper, temporally static VTA models are insufficient for capturing the influence of polarity. However, they can be useful to efficiently identify contacts closest to the target tract. Conversely, neuron models have the capacity to differentiate between various polarities, but they demand greater computational efforts. 





\section*{APPENDIX}

\section*{ACKNOWLEDGMENT}
The computations were enabled by resources provided by the National Academic Infrastructure for Supercomputing in Sweden (NAISS) and the Swedish National Infrastructure for Computing (SNIC) at UPPMAX partially funded by the Swedish Research Council through grant agreements no. 2022-06725 and no. 2018-05973.

Additionally, the authors
would like to thank Konstantin Butenko, Department of Neurology, Brigham and Women's Hospital, for his help
with OSS-DBS.
\bibliographystyle{IEEEtran}
\bibliography{IEEEabrv,references}
\addtolength{\textheight}{-12cm}

\end{document}